\documentclass[journal,twoside,onecolumn]{IEEEtran}

\usepackage{cite}
\usepackage{authblk}
\usepackage{setspace}
\usepackage{physics}
\usepackage{tikz}
\usetikzlibrary{quantikz2}
\usepackage{amssymb}
\usepackage{amsmath}
\usepackage{booktabs}
\usepackage{comment}
\usepackage{caption}
\usepackage{authblk}

\title{Bayesian optimization and nonlocal effects method for $\alpha$ decay of superheavy nuclei based on CPPM}
\author[1]{Xuanpeng Xiao}
\author[1]{Panpan Qi}
\author[1]{Gongming Yu$^{*}$\thanks{Corresponding author: \\ $^{*}$ ygmanan@kmu.edu.cn}}
\author[2]{Haitao Yang$^{\dag}$\thanks{$^{\dag}$yanghaitao205@163.com}}
\author[3]{Qiang Hu$^{\ddag}$\thanks{$^{\ddag}$qianghu@impcas.ac.cn}}

\affil[1]{College of Physics and Technology, Kunming University, Kunming 650214, China}
\affil[2]{College of Science, Zhaotong University, Zhaotong 657000, China}
\affil[3]{Institute of Modern Physics, Chinese Academy of Sciences, Lanzhou 730000, China}

\begin{document}

\maketitle

\begin{abstract}
We combine nonlocal effects with Bayesian Neural Network (BNN) methods to enhance the prediction accuracy of $\alpha$ decay half-lives. The results indicate that accounting for nonlocal effects significantly impacts the half-life calculations, while the BNN method markedly improves prediction accuracy and demonstrates strong extrapolation capabilities. Furthermore, we discuss the impact of nuclear deformation (the quadrupole deformation factor $\beta_2$) on machine learning predictions. Through Shapley Additive Explanations (SHAP), we conducted a quantitative comparison of six input features within the BNN, revealing that the $\alpha$ decay energy $Q_\alpha$ is the primary driving factor affecting the half-life $T_{1/2}$.  Leveraging the remarkable extrapolation ability of the BNN, we successfully predicted the $\alpha$ decay half-lives of the isotope chain ($Z=118, 120$), uncovering a significant shell effect at neutron number $N=184$. For the isotopic chains ($Z=118, 120$), the predicted $\alpha$ decay half-lives and $Q_{\alpha}$ values satisfy the Geiger-Nuttall (G-N) linear relationship. This result further confirms the predictive reliability of the proposed model.

\end{abstract}

\section{Introduction}
\label{chap:1} 
The synthesis and properties of superheavy nuclei represent a frontier in contemporary nuclear physics, playing a crucial role in exploring the limits of the nuclear chart and understanding fundamental nuclear forces. $\alpha$ decay is one of the primary decay modes of superheavy nuclei, accurate prediction of $\alpha$ decay half-lives is essential for identification and synthesis of new elements. $\alpha$ decay is fundamentally a quantum tunneling phenomenon, independently proposed by Gamow\cite{gamow1928quantentheorie} and Gurney and Condon\cite{gurney1928wave}\cite{condon1928wave} in 1928.

A variety of theoretical models have been developed to calculate the $\alpha$ decay half-lives of superheavy nuclei, including the Generalized Liquid Drop Model (GLDM)\cite{bao2014systematical}\cite{royer2001light} the Effective Liquid Drop Model (ELDM)\cite{cui2018alpha}, the Modified Generalized Liquid Drop Model (MGLDM) \cite{santhosh2018alpha}\cite{santhosh2019half}\cite{santhosh2020alpha}, the Coulomb and Proximity Potential Model (CPPM) \cite{santhosh2013role}, and the Preformed Cluster Model (PCM)\cite{singh2010cluster}. Additionally, empirical formulas based on fitting experimental data have been extensively developed. For example, the Universal Decay Law (UDL: Formula-A) proposed by Qi \cite{qi2009universal}, as well as the Modified Universal Decay Law that accounts for angular momentum effects (Formula-B) and isospin effects (Formula-C) \cite{soylu2021extended}. While these models provide essential theoretical foundations for understanding $\alpha$ decay mechanisms, significant discrepancies persist between theoretical predictions and experimental data, particularly in the superheavy region.

To improve prediction accuracy, researchers have focused on nonlocal effects in particle-nucleus interactions. The concept of nonlocal potentials was introduced by Feshbach\cite{feshbach1958optical} and Frahn et al.\cite{frahn1957velocity} in the late 1950s. Recent studies have further confirmed the importance of nonlocal effects: in 2022, Medeiros et al.\cite{medeiros2022nonlocality} conducted a systematic study of the $\alpha$ decay half-lives of $\alpha$ particles with effective mass for $52\leq Z\leq103$; Hu et al.\cite{hu2025nonlocality} systematically investigated nonlocal effects in $\alpha$ decay half-lives for even-even nuclei within a two potential approach. These studies consistently demonstrate that incorporating nonlocal effects significantly improves theoretical predictions, offering a new avenue for reducing systematic deviations between theory and experiment.

Traditional theoretical models face challenges in handling complex many-body interactions and nonlinear effects. Recent advances in machine learning have provided new tools for nuclear physics research. Machine learning exhibits unique advantages in treating high-dimensional, strongly correlated complex systems\cite{niu2019predictions}. Various machine learning methods have been successfully applied in nuclear physics, including Gaussian processes(GP)\cite{iwamoto2020generation}\cite{yuan2022theoretical} naive Bayes classifiers(NBC)\cite{ma2020predictions}\cite{liu2021improved} deep learning(DL)\cite{li2022deep}\cite{ma2023simple} restricted Boltzmann machines(RBM)\cite{melko2019restricted}\cite{vieijra2020restricted} radial basis function(RBF)\cite{niu2016improved}\cite{zheng2014mass} networks, Bayesian Neural Networks(BNN)\cite{xiao2025bayesian}\cite{neufcourt2019neutron}, and light gradient boosting machines(LightGBM)\cite{gao2021machine}. Among these, BNN have attracted particular attention due to their advantages in uncertainty quantification and overfitting prevention. By incorporating the prior distribution of parameters, and Bayesian inference, BNN not only provides accurate predictions but also quantifies prediction uncertainties\cite{niu2019predictions}. This capability has enabled BNN applications across multiple nuclear physics domains\cite{kononenko1989bayesian}\cite{lampinen2001bayesian}, including predictions of atomic radii\cite{utama2016nuclear}\cite{dong2022novel}, nuclear masses\cite{neufcourt2018bayesian}, and $\beta$ decay half-lives\cite{minato2022calculation}.

In this study, we systematically investigate the theoretical calculation and model optimization of $\alpha$ decay half-lives for superheavy nuclei. Initially, we calculated the $\alpha$ decay half-lives of 401 superheavy nuclides based on the CPPM framework. In the conventional CPPM, the $\alpha$ cluster preformation factor $P_{\alpha}$ is typically treated as a constant empirical value for specific types of parent nuclei. This simplified treatment stems from the smooth variation of $P_{\alpha}$ in open-shell regions~\cite{hodgeson2003cluster}; however, this assumption may be insufficient for accurately reproducing half-lives near shell closures. The preformation factor is a highly complex physical quantity that reflects the interplay of nuclear many-body correlations, shell effects, and cluster formation dynamics. Particularly near closed shells (such as  $Z=82$ and $N=126$), $P_{\alpha}$ exhibits significant variations\cite{andreyev2013signatures}\cite{deng2015realistic}. Recently, researchers have proposed a cluster formation model (CFM) based on $\alpha$-cluster formation energy~\cite{deng2015realistic}\cite{ahmed2013clusterization}\cite{ahmed2017alpha}, which extracts the $\alpha$ preformation factor by analyzing the binding energy differences of participating nuclides. This approach shows excellent agreement with various microscopic theoretical calculations. Therefore, we employ the CFM to systematically analyze separation energies and extract $P_{\alpha}$. 

To account for nonlocal dynamical effects, we introduce a coordinate-dependent $\alpha$-particle effective mass and recalculate the half-lives. The results demonstrate that the inclusion of nonlocal effects significantly impacts the predicted $\alpha$ decay half-lives, consistent with previous studies\cite{medeiros2022nonlocality}\cite{hu2025nonlocality}.
Subsequently, we construct a BNN model using mass number $A$, proton number $Z$, decay energy $Q_{\alpha}$, orbital angular momentum $l$, and preformation factor $P_{\alpha}$ as input parameters to optimize the CPPM, the nonlocally modified CPPM and three distinct parameterizations of the universal decay law (UDL: Formula-A, Formula-B, Formula-C). Given that prior research has shown that nuclear deformation can significantly alter the barrier shape and thereby affect the tunneling probability in $\alpha$ decay\cite{ni2010systematic,hassanzad2021theoretical}, we further incorporate the quadrupole deformation parameter $\beta_2$ in the BNN to improve prediction accuracy. By comparing the root-mean-square errors (RMSE) before and after optimization, we find that the BNN reduces the RMSE of all models by more than 23.25\%, with the CPPM achieving an improvement of 36.28\%. Using Shapley Additive Explanations (SHAP) for feature importance analysis, the results reveal that the decay energy $Q_{\alpha}$ plays a dominant role among the six input parameters.

Finally, we predict the $\alpha$ decay half-lives for isotope chains of $Z=118$ and $Z=120$. The predictions show that the logarithm of $\alpha$ half-life ($\log_{10}T$) exhibits the expected linear Geiger-Nuttall–type correlation with the decay energy $Q_{\alpha}$. This correlation is consistent with the relationship between $\log_{10}T$ and the negative logarithm of penetration probability ($-\ln$P). This demonstrates that our method not only reduces the RMSE but also preserves the fidelity of fundamental physical laws, providing compelling evidence that the approach captures the underlying physics.

This paper is organized as follows: Section 2 introduces $\alpha$ decay theoretical models and BNN methodology; Section 3 presents results and analysis; Section 4 is a summary.
    
\section{General Formalism}
\label{chap: Chapter 2} 
\subsection{\textbf{Coulomb and proximity potential model framework}}
The partial half-life is related to the decay constant $\lambda$ by\cite{santhosh2021cluster}
\begin{eqnarray}\label{y1}
T_{1/2}=\frac{\ln2}{\lambda}=\frac{\ln2}{\nu P_{\alpha}P},
\end{eqnarray}
where $\lambda$ is the decay constant, $\nu=1.0\times10^{22} s^{-1}$ is the frequency of the assault on the barrier\cite{poenaru1991cluster}\cite{poenaru2002systematics}\cite{poenaru2011single}. $P_{\alpha}$ refers to the preformation factor, which will be discussed in the following section.

The barrier penetration probability $P$ can be calculated within the semi-classical Wentzel-Kramers-Brillouin(WKB) approximation.
\begin{eqnarray}\label{y3}
P=\exp(-\frac{2}{\hbar}\int_{R_{in}}^{R_{out}}\sqrt{2\mu|V_{r}-Q_{c}|}\mathrm{d}r),
\end{eqnarray}
where the reduced mass is $\mu=\frac{mM}{m+M}$. Here, with $M$ and $m$ being the daughter nucleus and the emitted cluster mass, respectively. $Q_{c}$ represents the released energy\cite{saidi2015cluster}, which can be expressed as
\begin{eqnarray}\label{y4}
Q_{c}=B(A_{c},Z_{c})+B(A_{d},Z_{d})-B(A,Z),
\end{eqnarray}
where $B(A_{c}, Z_{c})$, $B(A_{d}, Z_{d})$ and $B(A, Z)$ are the binding energies of the emitted cluster, daughter nucleus, and parent nucleus, respectively\cite{yuan2022theoretical}. $Z_{c}$, $Z_{d}$ and $Z$ are the proton numbers of the emitted cluster, daughter nucleus, and parent nucleus, respectively. The classical turning points
$ R_{in}=A_{c}^{1/3}+A_{d}^{1/3}$ and $R_{out}=(\sqrt{(\frac{Z_{c}Z_{d}e^{2}}{2Q_{c}})^{2}+\frac{\hbar^{2}(l+1/2)^{2}}{2\mu Q_{c}}}+\frac{Z_{c}Z_{d}e^{2}}{2Q_{c}})$\cite{dong2009cluster}. Here, $A_{c}$, $A_{d}$, and $A$ are the mass numbers of the emitted cluster, the daughter nucleus, and the parent nucleus, respectively.

The total interaction potential $V(r)$ between the emitted cluster and the daughter nucleus consists of the nuclear potential $V_{N}(r)$, Coulomb potential $V_{C}(r)$, and centrifugal potential $V_{l}(r)$. It can be expressed as
\begin{eqnarray}\label{y5}
V(r)=V_{N}(r)+V_{C}(r)+V_{l}(r),
\end{eqnarray}

In the present work, we adopt the proximity potential formalism in place of the nuclear potential. In the 1970s, Blocki et al.\cite{blocki1977proximity} proposed the original form of the proximity potential formalism for two interacting spherical nuclei, which is expressed as
\begin{eqnarray}\label{y6}
V_{N}(r)=4\pi\gamma b\bar{R}\Phi(\xi),
\end{eqnarray}
where $b\approx1$fm. Surface energy coefficient $\gamma$ has the following form\cite{Myers:1967zza}:
\begin{eqnarray}\label{y7}
\gamma=\gamma_{0}(1-k_{s}I^{2}),
\end{eqnarray}
where $I=\frac{N_{p}-Z_{p}}{A_{p}}$ is the asymmetry parameter quantifying the neutron–proton excess of the parent nucleus; Here, $\gamma_{0}$ and $k_{s}$ denote the surface-energy coefficient and the surface-asymmetry coefficient, respectively. In the present work we use $\gamma_{0}$=0.9517MeV/fm$^{2}$ and $k_{s}=1.7826$\cite{Myers:1967zza}.

$\bar{R}$ denotes the mean curvature radius (or reduced radius). It can be obtained from
\begin{equation}\label{y8}
\bar{R}=\frac{C_{c}C_{d}}{C_{c}+C_{d}},
\end{equation}
where $C_{i}=R_{i}[1-(\frac{b}{R_{i}})^{2}](i=c, d)$ denotes the matter radius. The effective sharp radius $R_{i}$ is defined as $R_{i}=1.28A_{i}^{1/3}-0.76+0.8A_{i}^{-1/3}(\,i=c, d)$.

The general functional form is given by
\begin{eqnarray}\label{y9}
\Phi(\xi)\!=\!\!
\begin{cases}
\frac{-1}{2}(\xi-2.54)^{2}\!-\!0.0852(\xi-2.54)^{3},\xi<1.2511,\nonumber\\
-3.437\exp(-\frac{\xi}{0.75}),\quad\quad\quad\quad\quad\quad\xi\geq1.2511,
\end{cases}\\
\end{eqnarray}
where $\xi=\frac{r-C_{c}-C_{d}}{b}$ is the distance between the near surface of the emitted cluster and daughter nucleus.

Here, the potential of a uniformly charged sphere of radius $R$ is treated as the Coulomb potential, which is given by
\begin{equation}\label{y10}
V_{C} =
\begin{cases}
\frac{Z_{c}Z_{d}e^{2}}{2R}[3-(\frac{r}{R})^2],\quad\quad r\leq R,\\
\frac{Z_{c}Z_{d}e^{2}}{r},\quad\quad\quad\quad\quad\quad r\geq R,
\end{cases}
\end{equation}
where $e^2 = 1.43$ MeV$\cdot$ fm represents the Coulomb interaction constant, and $R$ is the Sharp radius, with $R_{c}$ and $R_{d}$ denoting the radii of the daughter nucleus and the emitted cluster, respectively.

In this work, we adopt $l(l+1)\rightarrow(l+1/2)^{2}$ as the Langer correction form, since it is a necessary correction for one-dimensional problems. It can be expressed as\cite{morehead1995asymptotics}
\begin{equation}\label{y11}
V_{l}(r)=\frac{(l+1/2)^{2}\hbar^{2}}{2\mu r^{2}},
\end{equation}
where $\hbar$ is the reduced Planck constant. $l$ is the angular momentum carried by the emitted cluster. It can be obtained by
\begin{equation}\label{y11}
l =
\begin{cases}
\bigtriangleup_{j}, & \text{for even } \bigtriangleup_{j} \text{ and } \pi_{p} = \pi_{d}, \\
\bigtriangleup_{j} + 1, & \text{for even } \bigtriangleup_{j} \text{ and } \pi_{p} \neq \pi_{d}, \\
\bigtriangleup_{j}, & \text{for odd } \bigtriangleup_{j} \text{ and } \pi_{p} \neq \pi_{d}, \\
\bigtriangleup_{j} + 1, & \text{for odd } \bigtriangleup_{j} \text{ and } \pi_{p} = \pi_{d}.
\end{cases}
\end{equation}
where $\pi_{p}$ and $\pi_{d}$ are the parity value of the parent nucleus and the daughter nucleus, respectively. $\bigtriangleup_{j}=|j_{p}-j_{d}-j_{c}|$, and $j_{p}$, $j_{d}$, $j_{c}$ are the isospin value of the parent nuclei, the daughter nuclei and the emitted cluster, respectively\cite{liu2023systematic}.

\subsection{Cluster formation model(CFM)}
The preformation factor $P_{\alpha}$, within the CFM, is expressible as as\cite{deng2015realistic} 
\begin{equation}\label{y12}
P_{\alpha}=\frac{E_{f\alpha}}{E},
\end{equation}
where the formation energy of the $\alpha$ cluster is designated as $E_{f\alpha}$, and the total energy of the system in question is referred to as E. The formation energy and total energy are determined by analyzing the neutron-neutron and proton-proton pairing interactions, along with the proton-neutron correlations. It is expressed as
\begin{equation}\label{y13}
\begin{aligned}
E_{ja} = & 3B(A,Z) + B(A-4,Z-2) \\
        & - 2B(A-1,Z-1) - 2B(A-1,Z)\\
        &\!\!\!\!\!\!= [2S_{p}(A,Z)+ 2S_{n}(A,Z)]-2S_{\alpha}(A,Z),
\end{aligned}
\end{equation}

\begin{equation}\label{y14}
E=S_{\alpha}(A,Z)=B(A,Z)-B(A-4,Z-2),
\end{equation}

The recently proposed CFM has been extended to evaluate the $\alpha$ preformation factors in odd-$A$ and odd-odd nuclei\cite{ahmed2013alpha}. Subsequently, Deng \textit{et al}. introduced an adaptive correction to the formation energy and investigated the $\alpha$ preformation in odd-$A$ and odd-odd superheavy systems, achieving good consistency with both microscopic expectations and experimentally extracted values. Moreover, the formation energy was recast in a unified form\cite{deng2015realistic}:
\begin{equation}\label{y15}
E_{f\alpha} = 
\begin{cases}
2S_p + 2S_n - S_\alpha & \text{(even-even)} \\
2S_p + S_{2n} - S_\alpha & \text{(even-odd)} \\
S_{2p} + 2S_n - S_\alpha & \text{(odd-even)} \\
S_{2p} + S_{2n} - S_\alpha & \text{(odd-odd)}
\end{cases}
\end{equation}
where $S_{p}(S_{n})$, $S_{2p}(S_{2n})$ are the one-proton (neutron) separation energy and two-proton (neutron) separation energy of the parent nucleus, respectively,
\begin{equation}\label{y16}
\begin{cases}
    S_p(A,Z)= B(A,Z) - B(A-1,Z-1)\\ 
    S_n(A,Z)= B(A,Z) - B(A-1,Z)\\
    S_{2p}(A,Z) = B(A,Z) - B(A-2,Z-2)\\
    S_{2n}(A,Z) = B(A,Z) - B(A-2,Z),
\end{cases}
\end{equation}
for the above formulas, we adopted existing experimental data and evaluated values from the Atomic Mass Evaluation (AME) table.

\begin{figure}[!ht]
    \centering
    \includegraphics[width=0.7\textwidth]{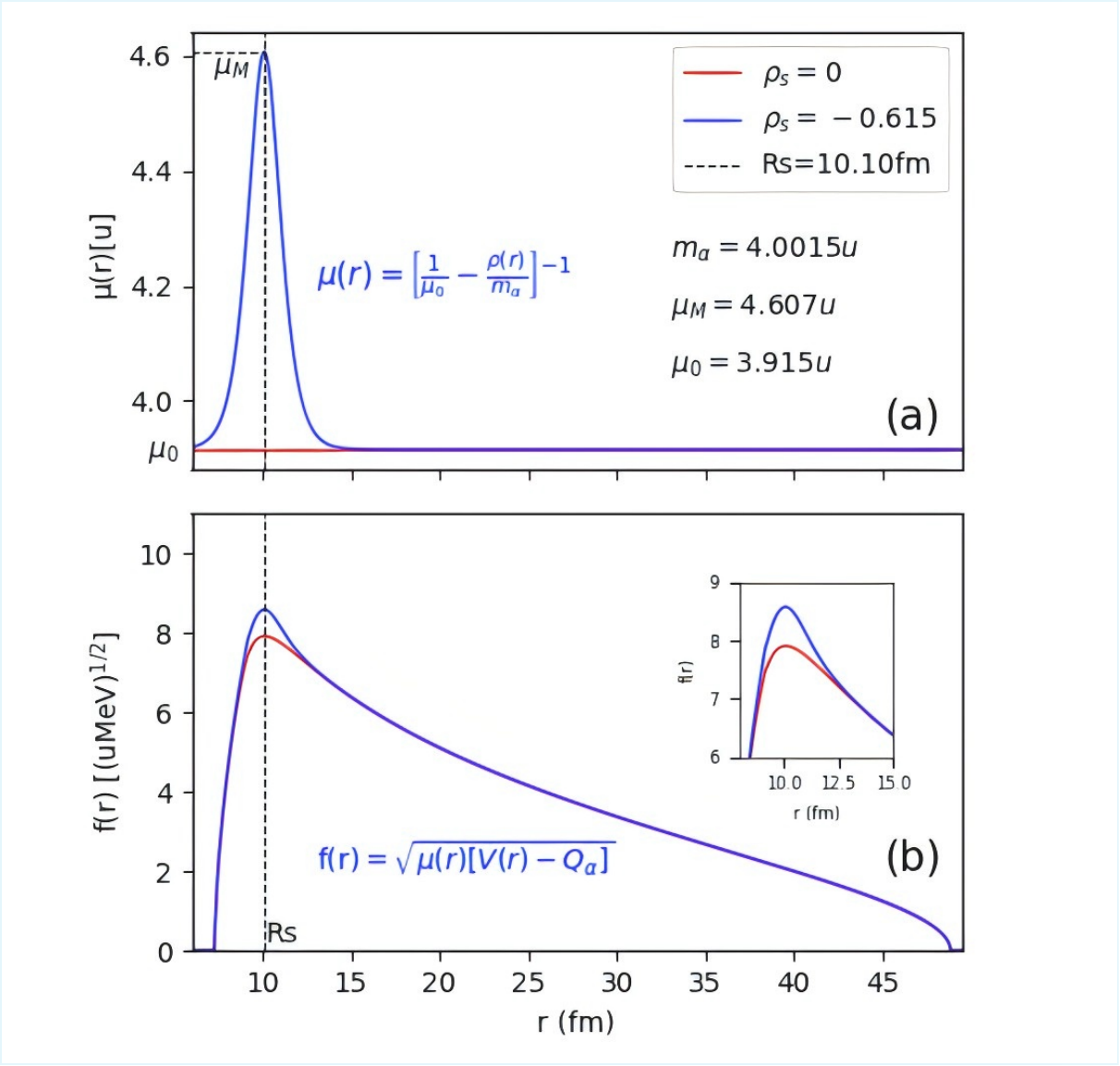}   
    \caption{The impact of nonlocal effects on tunneling calculations is illustrated using the example of $^{184m}_{78}Pt$ $\alpha$ decay. In panel (a), the effective reduced mass $\mu$ is presented considering nonlocal effects. Panel (b) compares the barrier penetration integral function $f(r)$: the reduced mass $\mu$ (blue curve, $\rho_{s} = -0.615$) versus $\mu_{0}$ (red curve, $\rho_{s} = 0$).}
    \label{fig:dataset}
\end{figure}

\subsection{Nonlocality effect}
In the present work, considering the nonlocal dynamical effects, a coordinate-dependent effective mass of the $\alpha$ particle is introduced, which can be expressed as
\begin{eqnarray}\label{y17}
\mu=\frac{m^*M}{m^*+M},
\end{eqnarray}
where M is the nuclear mass of the daughter nucleus. In this work, a coordinate-dependent effective mass m* correction is employed to describe the nonlocal dynamical effects in particle-nucleus interactions, which is expressed as the derivative of the Woods-Saxon function multiplied by a position-dependent effective mass function\cite{teruya2016nonlocality}\cite{jaghoub2011novel}\cite{zureikat2013surface}\cite{alameer2021nucleon}.
\begin{eqnarray}\label{y18}
m^*=\frac{m}{1-\rho(r)},
\end{eqnarray}
where m is the free mass of $\alpha$ particle and the $\rho(r)$ function is defined as
\begin{eqnarray}\label{y19}
\rho(r)=\rho_{s}a_{s}\frac{d}{dr}[1+\exp(\frac{r-R_{s}}{a_{s}})]^{-1},
\end{eqnarray}
where parameter $R_{s}$ is defined as $R_{s}=R+\Delta R$, representing the centroid position of the effective mass function $\rho(r)$, with $a_{s}$ related to the function width. The values $\Delta R=3.44$fm and $a_{s}=0.65$ are consistent with Ref. \cite{medeiros2022nonlocality}. Global fitting of the mass parameter $\rho_{s}$ to the complete experimental data set is necessary. Nonlocal effects can be incorporated into particle-nucleus interactions through energy-dependent potential terms\cite{liu2023systematic}. Within the WKB framework, the analytical form of the penetration factor P remains unchanged, requiring only the replacement of free mass m with effective mass m*. To illustrate the contribution of effective mass m* to tunneling calculations, Fig 1 presents the reduced effective mass $\mu(r)$ [Fig. 1(a)] and the function $f(r) = \sqrt{\mu(r)[V(r)-Q_{\alpha}]}$ in the barrier penetration probability P integral [Fig. 1(b)] for $\alpha$ decay of $^{184m}_{78}Pt$. The results reveal that nonlocality significantly affects the reduced effective mass $\mu(r)$ near the nuclear surface, thereby modifying the profile of the penetration integral function compared to free-mass calculations.

\begin{table}
\centering
\caption{RMSE deviations $\sigma_{\text{pre}}$ of $\alpha$ decay half-lives $T_{1/2}$ calculated by the original CPPM, CPPM with nonlocalization considered, and three empirical formulas (UDL: Formula A, B, and C), as well as the BNN corrected $\sigma_{\text{post}}$ (based on a complete dataset containing 401 $\alpha$ decay cases)}
\renewcommand{\arraystretch}{1.4}
\begin{tabular}{c|ccc}
\hline
\hline
   Models &\quad\quad $\sigma_{pre}$ &\quad\quad $\sigma_{post}$ &\quad\quad $\Delta\sigma/\sigma_{post}$     \\
\hline
CPPM &\quad\quad $0.7268$ &\quad\quad $0.4631$ &\quad\quad $36.28\%$     \\
CPPM(nonlocality)&\quad\quad$0.4946$&\quad\quad$0.3796$&\quad\quad $23.25\%$     \\
UDL(Formula-A) &\quad\quad $0.6348$ &\quad\quad $0.4121$ &\quad\quad $35.99\%$     \\
UDL(Formula-B) &\quad\quad $0.6342$ &\quad\quad $0.4345$ &\quad\quad $31.49\%$     \\
UDL(Formula-C)&\quad\quad $0.7206$ &\quad\quad $0.4596$ &\quad\quad $36.22\%$     \\
\hline
\hline
\end{tabular}
\label{Tensor D}
\end{table}

\subsection{Empirical formula}
\subsubsection{Formula-A: The UDL formula without any modifications that was obtained by Qi et al. is expressed as\cite{qi2009universal}}
\begin{eqnarray}\label{y16}
\log_{10}(T_{1/2})=aZ_{c}Z_{d}\sqrt{\frac{U}{Q_{c}}}+b\sqrt{UZ_{c}Z_{d}(A_{d}^{1/3}+A_{c}^{1/3})}+c,
\end{eqnarray}
where $U=\frac{A_{c}A_{d}}{A_{c}+A_{d}}$ and the adjustable parameters are $a=0.4314$, $b=-0.4087$, and $c=-25.7725$\cite{qi2009universal}.

\subsubsection{Formula-B: The UDL formula for angular momentum cases, as proposed by Qi et al., takes the following form\cite{qi2012effects}}
\begin{eqnarray}\label{y17}
\log_{10}(T_{1/2})=aZ_{c}Z_{d}\sqrt{\frac{U}{Q_{c}}}+b\sqrt{UZ_{c}Z_{d}(A_{d}^{1/3}+A_{c}^{1/3})}\nonumber\\
+c+d\sqrt{UZ_{c}Z_{d}(A_{d}^{1/3}+A_{c}^{1/3})}\sqrt{l(l+1)},
\end{eqnarray}
where the parameters $a$, $b$, and $c$ adopt the same values as in the UDL version, while the parameter $d=0.00476$ is taken from Ref. \cite{soylu2021extended}.

\subsubsection{Formula-C: The UDL formula, modified to incorporate the isospin effect, can be expressed as follows}
\begin{eqnarray}\label{y18}
\log_{10}(T_{1/2})=aZ_{c}Z_{d}\sqrt{\frac{U}{Q_{c}}}+b\sqrt{UZ_{c}Z_{d}(A_{d}^{1/3}+A_{c}^{1/3})}\nonumber\\
+c+e\sqrt{I(I+1)},
\end{eqnarray}
where $I=\frac{A-2Z}{A}$. The parameter $a, b, c, e$ are taken from Ref. \cite{soylu2021extended}.

\begin{table}[!ht]
\centering
\caption{The RMSE deviation of the half-life $T_{1/2}$ provided by the theoretical model is denoted as $\sigma_{\text{pre}}$, while the RMSE deviation following calibration with the BNN is denoted as$\sigma_{\text{post}}$. A total of 401 data sets were randomly divided into a training set ($80\%$, 300 sets) and a validation set ($20\%$, 101 sets).}
\renewcommand{\arraystretch}{1.4}
\begin{tabular}{c|cccccc}
\hline
\hline
        Models & \multicolumn{3}{c}{Learning set} & \multicolumn{3}{c}{Validation set} \\
        \cline{2-7}
        & $\sigma_{\text{pre}}$ & $\sigma_{\text{post}}$ & $\Delta\sigma/\sigma_{\text{pre}}$&\quad$\sigma_{\text{pre}}$&$\sigma_{\text{post}}$&$\Delta\sigma/\sigma_{post}$ \\
        \hline
        $CPPM$&0.7178& $0.4283$ & $40.41\%$&\quad$0.7580$&$0.5509$&$27.32\%$ \\
        $CMMP(nonlocality)$&0.4924& $0.3738$ & $24.09\%$&\quad$0.5032$&$0.4398$&$12.60\%$ \\
        $UDL(Formula-A)$ &0.6352& $0.4044$ & $36.34\%$&\quad$0.6768$&$0.4345$&$35.80\%$ \\
        $UDL(Formula-B)$&0.6287& $0.4166$ & $33.74\%$&\quad$0.6555$&$0.4906$&$25.16\%$ \\
        $UDL(Formula-C)$&0.7131& $0.4486$ & $37.09\%$&\quad$0.7496$&$0.4880$&$34.90\%$ \\
\hline
\hline
\end{tabular}
\end{table}

\subsection{Bayesian Neural Network}
BNN constitute a probabilistic architecture, with comprehensive details provided in Ref.\cite{neal2012bayesian}; This work focuses solely on their essential characteristics. Within the Bayesian framework, model parameters $\omega$ are represented as probability distributions rather than the deterministic values employed in conventional neural networks. A prior distribution $p(\omega)$ is introduced over all possible values of $\omega$. Given a dataset $D = \{(x_i, t_i) \mid i = 1, 2, \ldots, n\}$, where $x_{i}$ and $t_{i}$ denote the input and target output samples respectively, and $n$ represents the sample size, the posterior distribution $p(\omega|D)$ can be derived through Bayes theorem upon incorporation of the data $D$.
\begin{eqnarray}\label{y19}
p(\omega|D)=\frac{p(D|\omega)p(\omega)}{p(D)},
\end{eqnarray}
where $p(D|\omega)$ is the likelihood function, $p(D)$ is a normalization constant, which ensures the posterior distribution is a valid probability density and integrates to one.

\begin{table}[!ht]
\centering
\caption{Effect of different feature inputs on the RMSE deviation of $\alpha$ decay half-life $T_{1/2}$ calculated by various models using the BNN Method. The quadrupole deformation parameters $\beta_{2}$ are taken from Ref.\cite{moller1993nuclear}}
\renewcommand{\arraystretch}{1.4}
\begin{tabular}{c|c|c}
\hline
\hline
Models&Inputs&RMSE\\
\hline
CPPM&$A, Z, Q, l, Pa$&0.4631\\
CPPM&$A, Z, Q, l, Pa, \beta_{2}$&0.4535\\
\hline
CMMP(nonlocality)&$A, Z , Q, l, Pa$&0.3796\\
CMMP(nonlocality)&$A, Z, Q, l, Pa, \beta_{2}$&0.3733\\
\hline
UDL(Formula-A)&$A, Z, Q, l, Pa$&0.4121\\
UDL(Formula-A)&$A, Z, Q, l, Pa, \beta_{2}$&0.4113\\
\hline
UDL(Formula-B)&$A, Z, Q, l, Pa$&0.4345\\
UDL(Formula-B)&$A, Z, Q, l, Pa, \beta_{2}$&0.4293\\
\hline
UDL(Formula-C)&$A, Z, Q, l, Pa$&0.4596\\
UDL(Formula-C)&$A, Z, Q, l, Pa, \beta_{2}$&0.4373\\
\hline
\hline
\end{tabular}
\end{table}
In this study, the prior distribution $p(\omega)$ is modeled as a zero-mean Gaussian, whose precision (inverse variance) is governed by a gamma distribution. This configuration allows the precision parameter to vary over a wide range, enabling the BNN method to automatically determine its optimal value during sampling\cite{niu2018nuclear}. The likelihood function is commonly modeled as a Gaussian distribution, $p(D|\omega)=\exp(-\chi^2/2)$, where
\begin{eqnarray}\label{y20}
\chi^2=\sum^{N}_{n=1}(\frac{t_{n}-S(x;\omega)}{\Delta t_{n}})^2,
\end{eqnarray}
here, $\Delta t_{n}$ denotes the associated noise error for the i-th observable, and N represents the total number of observational data points. In the BNN framework, the network $S(x;\omega)$ can be expressed as:
\begin{eqnarray}\label{y21}
S(x;\omega)=a+\sum^{H}_{j=1}b_{j}tanh(c_{j}+\sum^{I}_{i=1}d_{ji}x_{i}),
\end{eqnarray}
where $x=\{x_{i}\}$ represents the input data, $w=\{a,b_{j},c_{j},d_{ji}\} $ are the model free parameters. $H$ and $I$ denote, respectively, the number of hidden-layer neurons and the dimensionality of the input. Given the prior distribution $p(\omega)$ and likelihood function $p(D|\omega)$, the posterior distribution $p(\omega|D)$ is obtained via variational inference to compute BNN predictions:
\begin{eqnarray}\label{y22}
\langle S\rangle=\int S(x;\omega)p(\omega|D)d\omega,
\end{eqnarray}

Within the Bayesian inference framework, variational inference approximates the intractable posterior distribution $p(\omega|D)$ by introducing a parameterized distribution $q_{\theta}(\omega)$, where $\theta$ denotes the variational parameters\cite{mostafa2023review}. The core objective is to minimize the Kullback-Leibler(KL) divergence\cite{kullback1951information} between the variational and true posterior distributions:
\begin{eqnarray}\label{y23}
KL(q_{\theta}(\omega)||p(\omega|D))=\int q_{\theta}(\omega)\log(\frac{q_{\theta}(\omega)}{p(\omega|D)})d\omega,
\end{eqnarray}

\begin{table}[!ht]
\centering
\caption{Based on the CPPM: comparison of optimization effects on the CPPM via different strategies (considering local effects, BNN Method, and BNN method combined with nonlocal).}
\renewcommand{\arraystretch}{1.4}
\begin{tabular}{c|cc}
\hline
\hline
Method&$\sigma_{post}$ &$\Delta\sigma/\sigma_{post}$\\
\hline
nonlocality&0.4946&$31.95\%$\\
\hline
BNN&0.4631&$36.28\%$\\
\hline
nonlocality+BNN&0.3796&$47.77\%$\\
\hline
\hline
\end{tabular}
\end{table}

Since direct computation of the posterior $p(\omega|D)$ is typically intractable in practice, we introduce the evidence lower bound (ELBO) as an alternative optimization objective:
\begin{eqnarray}\label{y24}
KL(q_{\theta}||p) &=& \log p(D) - \int q_{\theta}(\omega)\log p(D|\omega)\,d\omega - \int q_{\theta}(\omega)\log\left(\frac{q_{\theta}(\omega)}{p(\omega)}\right)\,d\omega\nonumber\\
&=& \log p(D) - \left[E_{q_{\theta}(\omega)}[\log p(D|\omega)] - KL(q_{\theta}(\omega)||p(\omega))\right] \nonumber\\
&=& \log p(D) - \mathcal{L}(\theta)
\end{eqnarray}
where
\begin{eqnarray}\label{y25}
\mathcal{L}(\theta)=E_{q_{\theta}(\omega)}[\log(p(D|\omega))-KL(q_{\theta}(\omega)||p(\omega))],
\end{eqnarray}

The first term represents the expected log-likelihood, encouraging the variational distribution to better fit the observed data, while the second term serves as a KL regularization that constrains the variational distribution to remain close to the prior $p(\omega)$. Since $\log(P(\omega))$ is constant with respect to $\theta$, maximizing the ELBO is equivalent to minimizing the original KL divergence. The optimization is performed using the Bayes by Backprop algorithm\cite{blundell2015weight}, which employs Monte Carlo sampling to obtain unbiased gradient estimates for efficient parameter updates.

In this work, BNN is employed to directly train the residuals $t_{k}$, establishing their implicit correlations with the characteristic parameters $A$, $Z$, $Q_{c}$, $l$, $P_{\alpha}$ and $\beta_{2}$. Here, $t_{k}=\log_{10}(T_{1/2}^{\exp})-\log_{10}(T_{1/2}^{th})=\log_{10}({T_{1/2}^{\exp}/(T_{1/2}^{th}})$. The RMSE deviation $\sigma$ is used to assess the predictive accuracy of BNN corrected models:
\begin{eqnarray}\label{y26}
\sigma=\sqrt{\frac{1}{n}\sum^{n}_{i=1}[\log_{10}(\frac{T_{1/2}^{\exp}}{T_{1/2}^{th}})]^2},
\end{eqnarray}
where, $\log_{10}T_{1/2}^{\exp}$ represents the experimental half-life of nuclear decay, while $\log_{10}T_{1/2}^{th}$ denotes the theoretical half-life calculated by the model.

\begin{figure}[!ht]
    \centering
    \includegraphics[width=0.7\textwidth]{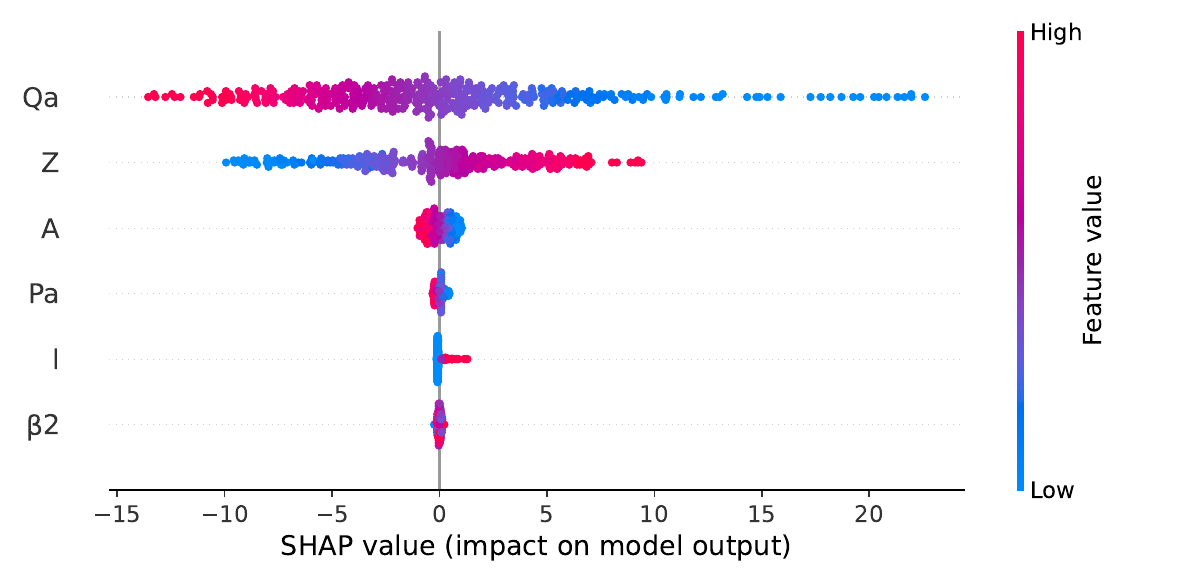}   
    \caption{The figure displays the importance ranking of four input features obtained using the SHAP toolkit. Each row represents a feature, with the horizontal axis showing its SHAP value, which reflects the feature's significance in the specific prediction. Each point corresponds to a sample, and the color of the points indicates the feature value, with red representing high values and blue representing low values.}
    \label{fig:dataset}
\end{figure}

\section{Results and discussion}
In this study, we utilize a total of 401 datasets, which include the experimental decay half-lives $\log_{10}T_{1/2}^{\exp}$, the angular momentum $l$ of the emitted cluster, $\alpha$ decay energy $Q_{c}$ and the parameters required to calculate $P_{\alpha}$ using the CFM method, with all data sourced from the Atomic Mass Evaluation (AME) table\footnote{{http://www.nndc.bnl.gov/ensdf}}. Recent research by Medeiros demonstrates significant improvements in the theoretical calculations of the $\alpha$ decay half-lives for even-even nuclei using the semiclassical WKB method, achieved through the introduction of coordinate-dependent effective mass. Furthermore, Hu et al. systematically investigated nonlocal effects in the $\alpha$ decay half-lives of even-even nuclei within the framework of a two-potential model. The results indicate a significant improvement in model accuracy following the introduction of the effective mass parameter for the $\alpha$ particle.

To further investigate the application of effective mass in theoretical models of $\alpha$ decay half-lives, We incorporate a coordinate-dependent effective mass parameter into the CPPM. We calculate the $\alpha$ decay half-lives for 401 nuclei and compare these results with those obtained from CPPM without considering effective mass. In the calculations of $\alpha$ decay half-lives within the dataset, the parameter $\rho_{s}$ is a critical parameter that is adjusted to minimize the RMSE deviations. Comparing with experimental results, when the coordinate-dependent effective mass is not considered (i.e., with the adjustment parameter $\rho_{s}=0$), the standard deviation is $\sigma_{\rho_{s}=0}=0.7268$. However, when the parameter $\rho_{s}$ is adjusted to $-0.615$, the standard deviation reduces to $\sigma_{\rho_{s}=-0.615}=0.4946$. These results indicate a significant enhancement in the CPPM calculations when incorporating the coordinate-dependent effective mass (with $\rho_{s}=-0.615$). The RMSE deviations decreased by approximately $31.95\%$. We summarize the effects of the potential nonlocal dynamic effects. Fig 1 illustrates the reduced effective mass $\mu$ during the $\alpha$ decay of $^{184m}_{78}Pt$ (Fig.1(a)) and its simplified form of the potential function $f(r) = \sqrt{\mu(r)[V(r)-Q_{\alpha}]}$ (Fig.1(b)). As shown in Fig.1(a), when considering the nonlocality of the potential, the effective mass decreases significantly; specifically, at $r=R_{s}$, the effective mass is reduced by approximately $17.7\%$, with a corresponding change in the simplified form of the potential function (as shown in Fig.1(b)). These results are consistent with the conclusions of Medeiros et al., further demonstrating the significant impact of nonlocal effects on $\alpha$ decay half-lives. Furthermore, accounting for these effects enhances the accuracy of the CPPM predictions for $\alpha$ decay half-lives.
\begin{table}
\centering
\caption{Predicted $\alpha$ decay half-lives of 14 nuclei ($Z=118, 120$) using BNN in logarithmic form. The values of $Q_{\alpha}^{RCHB}$, $Q_{\alpha}^{WS4}$, and $Q_{\alpha}^{FRDM}$ denote data derived from the RCHB\cite{wang2014surface}, WS4\cite{xia2018limits}, and FRDM\cite{moller2016nuclear} mass tables, respectively, with units in MeV.}
\begin{tabular}{c|cc|cc|cc}
\hline
 \hline
        $nucleus$ & $Q_{\alpha}^{RCHB}$ & $\log_{1/2}^{RCHB}$ & $Q_{\alpha}^{WS4}$&$\log_{1/2}^{WS4}$&$Q_{\alpha}^{FRDM}$&$\log_{1/2}^{FRDM}$ \\
        \hline
        $^{292}118$&10.96& $-0.880$ & $12.24$&\quad$-3.803$&$12.38$&$-4.110$ \\
        $^{294}118$&10.91& $-0.813$ & $12.20$&\quad$-3.784$&$12.36$&$-4.138$ \\
        $^{296}118$&10.77& $-0.534$ & $11.75$&\quad$-2.831$&$12.27$&$-4.010$ \\
        $^{298}118$&10.61& $-0.202$ & $12.18$&\quad$-3.880$&$12.48$&$-4.548$ \\
        $^{300}118$&10.47& $0.091$  & $11.95$&\quad$-3.425$&$12.50$&$-4.666$ \\
        $^{302}118$&10.61& $-0.303$ & $12.04$&\quad$-3.700$&$12.61$&$-4.985$ \\
        $^{304}118$&12.65& $-5.150$ & $13.12$&\quad$-6.204$&$13.38$&$-6.793$ \\
        $^{296}120$&11.87& $-2.413$ & $13.34$&\quad$-5.570$&$13.58$&$-6.086$ \\
        $^{298}120$&11.76& $-2.225$ & $12.91$&\quad$-4.734$&$13.23$&$-5.412$ \\
        $^{300}120$&11.62& $-1.967$ & $13.32$&\quad$-5.679$&$13.68$&$-6.460$ \\
        $^{302}120$&11.51& $-1.775$ & $12.89$&\quad$-4.837$&$13.55$&$-6.254$ \\
        $^{304}120$&11.72& $-2.316$ & $12.76$&\quad$-4.631$&$13.54$&$-6.310$ \\
        $^{306}120$&13.58& $-6.477$ & $13.78$&\quad$-6.919$&$14.26$&$-7.977$ \\
        $^{308}120$&13.07& $-5.443$ & $12.96$&\quad$-5.206$&$12.96$&$-5.206$ \\
        \hline
\hline
\end{tabular}
\end{table}

In this section, we comprehensively evaluate the global optimization performance and extrapolation capability of the BNN approach. Five benchmark models are selected for comparative analysis: the CPPM, the nonlocally modified CPPM, and three distinct parameterizations of the UDL(Formula-A, Formula-B, Formula-C). The complete dataset comprising 401 $\alpha$ decay nuclides is employed for model training and validation. We first calculate the raw residuals $t_k(A, Z, Q_{\alpha}, \ell, P_{\alpha})$ for each benchmark model and determine the RMSE $\sigma_{\mathrm{pre}}$ across the entire dataset; detailed numerical values are presented in Table~1. Subsequently, using the raw residuals $t_k(A, Z, Q_{\alpha}, \ell, P_{\alpha})$ as input features, we systematically calibrate the theoretical predictions through the BNN method, yielding corrected half-lives $T_{1/2}$. To quantitatively assess the improvement achieved by BNN calibration, Table~1 also reports the relative improvement ratio 
\begin{equation}
\frac{\Delta\sigma}{\sigma_{\mathrm{pre}}} = \frac{\sigma_{\mathrm{pre}} - \sigma_{\mathrm{post}}}{\sigma_{\mathrm{pre}}},
\end{equation}
where $\sigma_{\mathrm{post}}$ denotes the RMSE after calibration.

From Table 1, it is clear that although the CPPM exhibits excellent robustness, there remain notable discrepancies between the theoretical $T_{1/2}$ and the experimental data. Following BNN calibration, the prediction accuracy for $\alpha$ decay based on CPPM(nonlocal) has increased by over $23.25\%$. This section presents a systematic assessment of the global optimization capability and extrapolation performance of the BNN approach. Using the relative RMSE reduction $\Delta\sigma/\sigma_{\mathrm{pre}}$ the predictive performance of the three distinct parameterizations of the UDL(Formula-A, Formula-B, Formula-C) has improved by 35.99\%, 31.49\%, and 36.22\%, respectively. This underscores the ability of the BNN method to effectively uncover latent half-life correlations among different nuclei, thereby optimizing computational results and significantly enhancing the predictive capability of model-based semi-empirical formulas. A comparison of the data in Table 1 leads to the conclusion that the BNN method is applicable to both CPPM and various empirical formulas, with accuracy improvements reliant on the judicious selection of feature inputs.
\begin{figure}[!ht]
    \centering
    \includegraphics[width=0.7\textwidth]{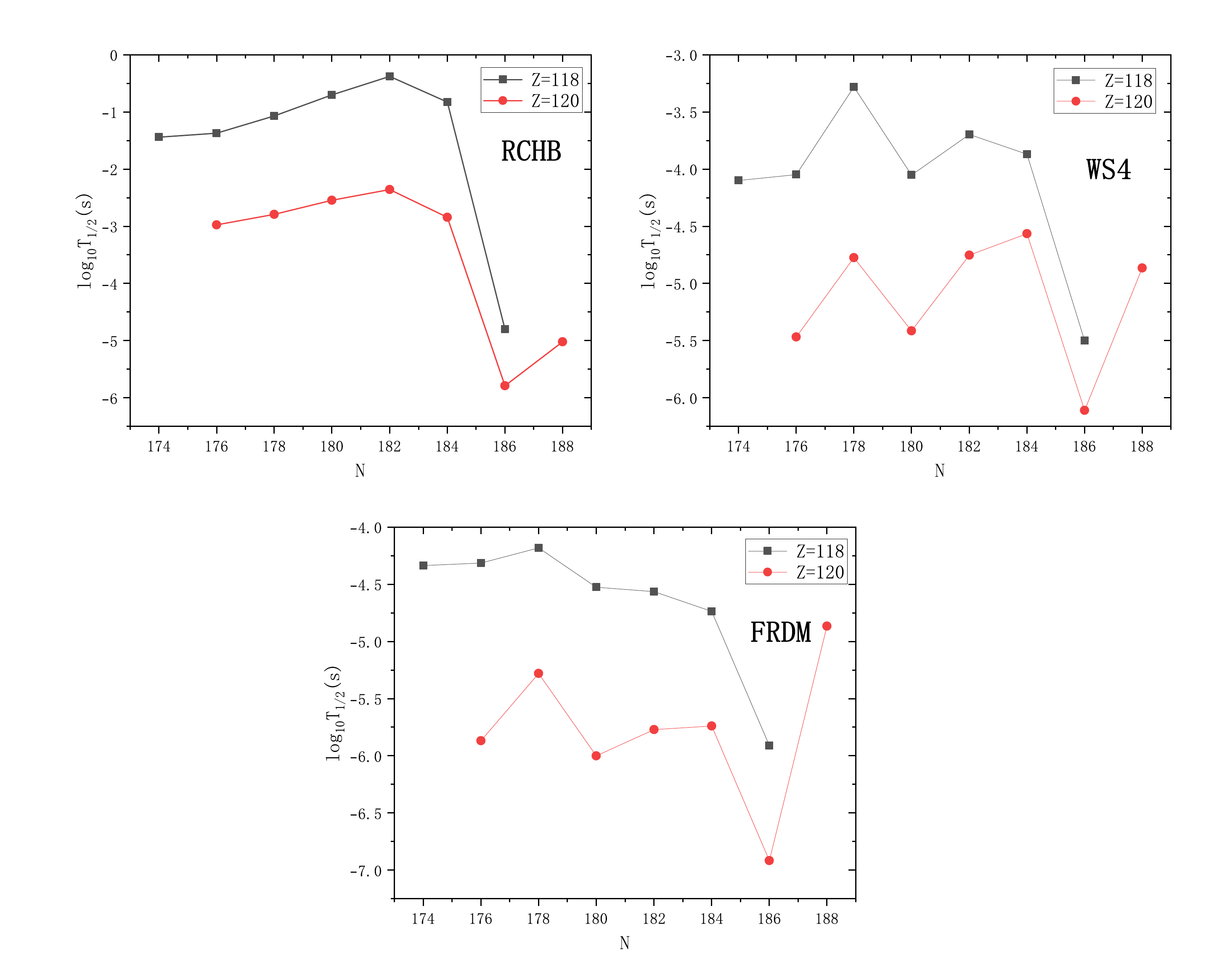}   
    \caption{$\alpha$ decay half-lives of isotopes with atomic numbers $Z=118$ and $Z=120$ predicted using the BNN method. Herein, black squares and red dots correspond to $Z = 118$ and $Z = 120$, respectively; the horizontal axis denotes the neutron number $N$, and the vertical axis represents the logarithm of the half-life, $\log_{10}T_{1/2}$; three mass tables (RCHB, WS4, FRDM) are employed in this figure.}
    \label{fig:dataset}
\end{figure}

\begin{figure}
    \centering
    \includegraphics[width=0.7\textwidth]{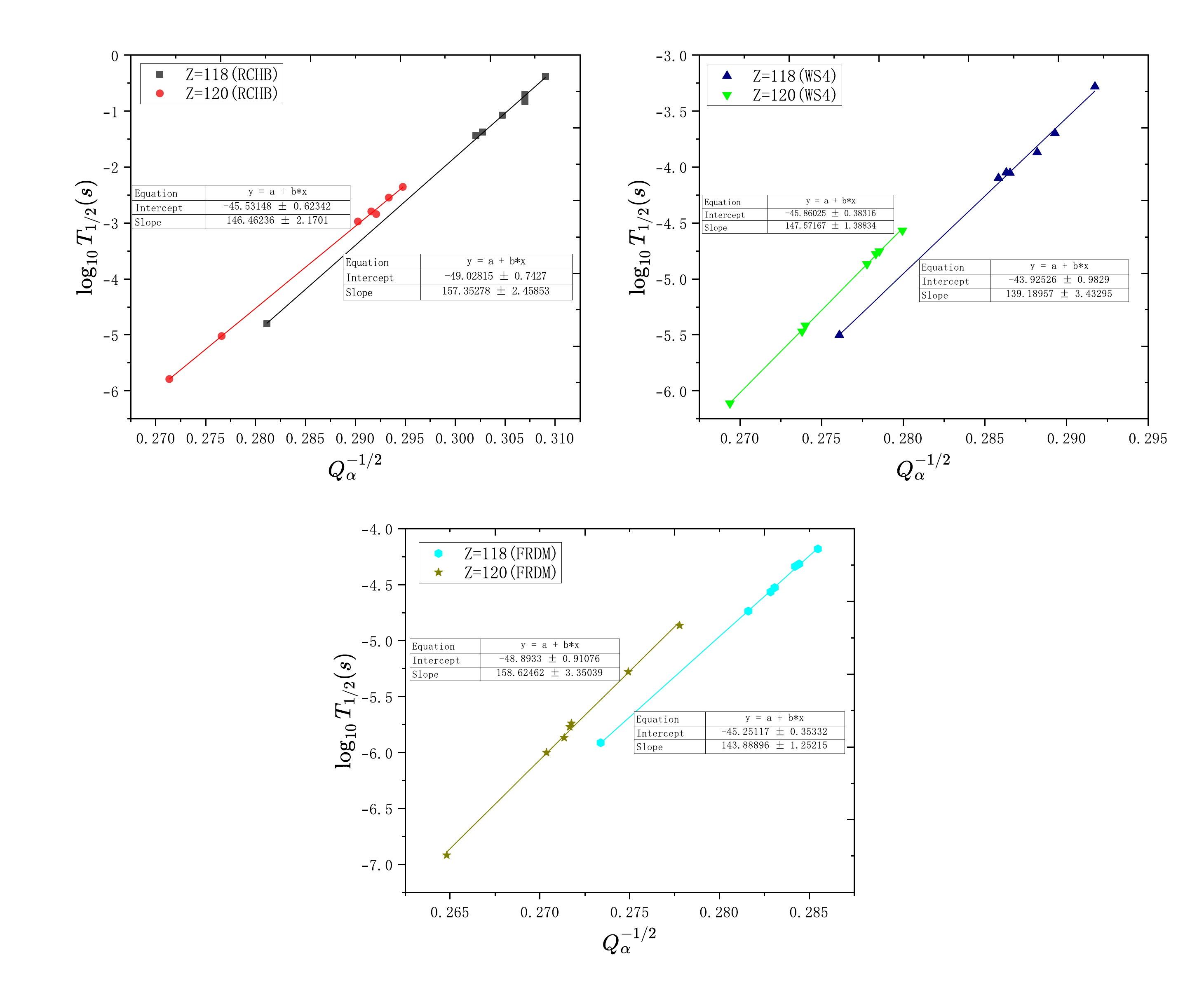}   
    \caption{G-N plots of $\alpha$ decay for Z=118 and Z=120 isotopes predicted by the BNN method.}
    \label{fig:dataset}
\end{figure}

In the field of nuclear physics, the prediction of nuclear decay half-lives holds significant practical value; however, it faces considerable challenges, particularly in regions with scarce experimental data. This section focuses on evaluating the capability of the BNN method for extrapolating $\alpha$ decay half-life predictions. Prior to the predictions, we randomly divided 401 data sets into a training set ($80\%$, 300 sets) and a validation set ($20\%$, 101 sets). Initially, we determined the neural network parameters using the training set data, and upon completing the model calibration, we calculated the RMSE deviations and improvement rates $\Delta\sigma/\sigma_{post}$ for both the training and validation sets, as detailed in Table 2. After posterior correction based on the BNN, compared with the uncorrected original CPPM, the RMSE of the model decreases by 40.41\% on the training set and by 27.32\% on the validation set. For the nonlocally modified CPPM, the corresponding RMSE reductions are 24.09\% and 12.60\%. Across the three empirical formulas examined, the BNN likewise yields substantial improvements, particularly for Formula A and Formula B, for which the training and validation set RMSEs are both reduced by more than 34.90\%. Altogether, results based on CPPM and multiple $\alpha$ decay empirical relations indicate that the BNN provides reliable extrapolation in half-life prediction and delivers highly credible estimates.

Building on the preceding calculations, we incorporate the quadrupole deformation parameter $\beta_2$ into the feature vector $(A, Z, Q_{\alpha}, \ell, P_{\alpha}, \beta_2)$ and re-calibrate the BNN for each physical model and empirical relation. The results are summarized in Table~3. Relative to the deformation-agnostic BNN, including $\beta_2$ yields additional RMSE reductions of 0.0096, 0.0063, 0.0008, 0.0052, and 0.0023 for CPPM, CPPM (nonlocal), and three distinct parameterizations of the UDL(Formula-A, Formula-B, Formula-C), respectively. These findings are consistent with Ref.~\cite{you2024calculating}.

Table~4 provides a clear comparison of the optimization gains achieved for CPPM by different strategies. When considering only localization effects or when directly optimizing CPPM with the BNN, the relative reduction in RMSE $\Delta\sigma/\sigma_{\mathrm{pre}}$ reaches 31.95\% and 36.28\%, respectively. By contrast, applying BNN optimization to the nonlocal CPPM yields $\Delta\sigma/\sigma_{\mathrm{pre}}= 47.77\%$, indicating a markedly stronger improvement and highlighting the synergy between nonlocality and BNN modeling in enhancing CPPM's predictive accuracy.

The interpretability of machine learning has emerged as a key area of research. Through interpretability analysis, we gain deeper insights into how algorithms extract meaningful information from vast datasets. To unveil the learning patterns of the CPPM, this study employs Shapley Additive Explanations to calculate SHAP values for each signal sample, thereby identifying the features that most significantly impact the prediction of T values. The arrangement of signals reflects their contribution levels. Each row represents a signal, with red points indicating higher value data points and blue points indicating lower value data points. Points on the right indicate a significant positive impact of the feature on the prediction, while points on the left suggest a negative influence. Fig 2 presents the importance ranking of six features, revealing that the decay energy $Q_{\alpha}$ and the proton number $Z$ are crucial driving factors in the CPPM predictions. This finding provides significant insights into the understanding of nuclear decay mechanisms.

\begin{figure}[!ht]
    \centering
    \includegraphics[width=0.7\textwidth]{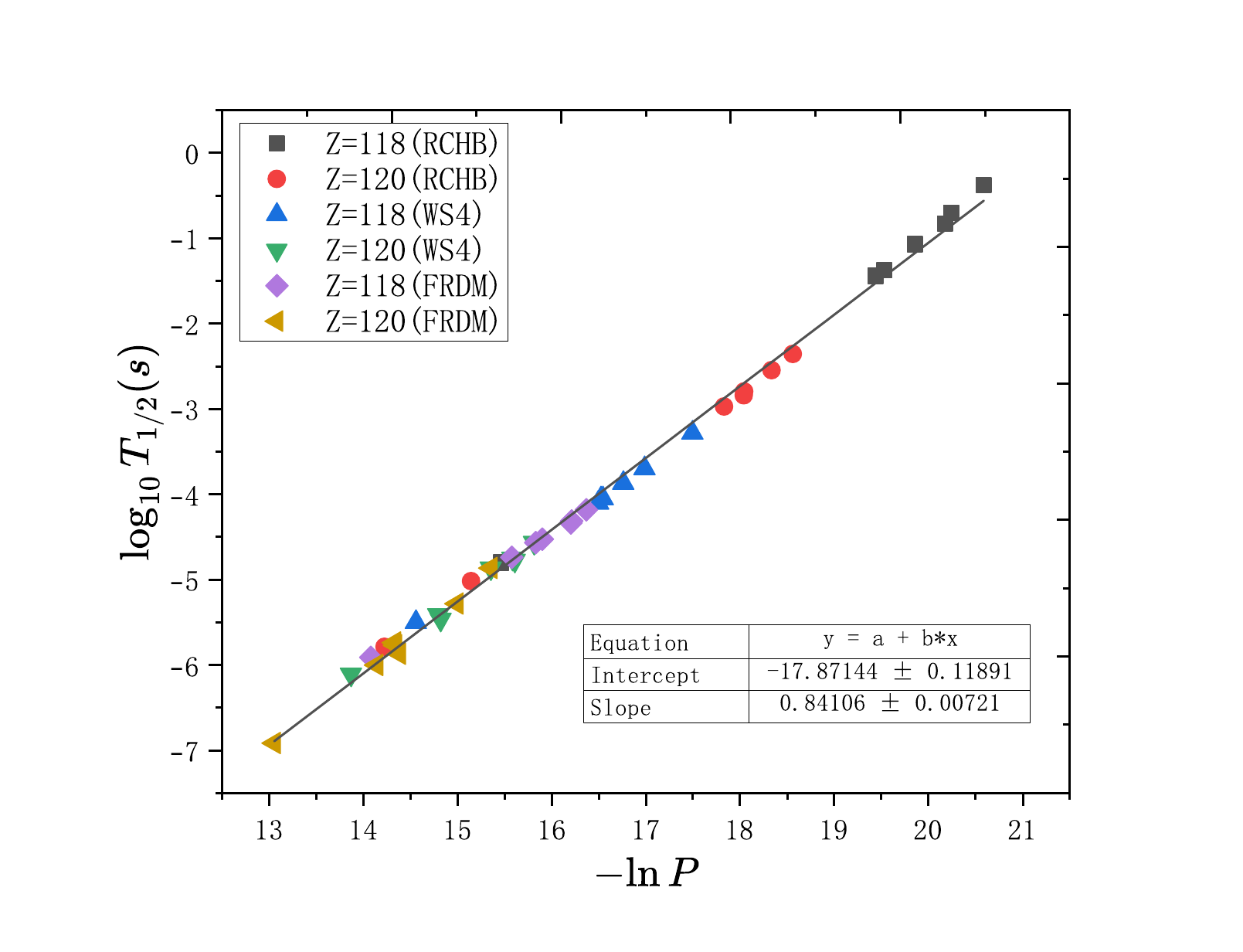}   
    \caption{Universal curves of $\alpha$ decay half-lives for isotopes with atomic numbers $Z=118$ and $Z=120$ versus negative logarithm of penetrability ($-\ln$P) predicted using BNN method.}
    \label{fig:dataset}
\end{figure}

This study employs BNN approach to predict the $\alpha$ decay half-lives. Given the high sensitivity of half-lives to the $\alpha$ decay energy ($Q_{\alpha}$ value), selecting appropriate mass models for calculation is crucial. For comparative purposes, we utilized three mass models: Relativistic Continuum Hartree-Bogoliubov (RCHB)\cite{wang2014surface}, Weizsacker–Skyrme-4(WS4)\cite{xia2018limits}, and Finite Range Droplet Model (FRDM)\cite{moller2016nuclear} to compute $Q_{\alpha}$ values. The calculation results are presented in Table 5, where the first column lists the $\alpha$ decay parent nuclei, The second to third columns, fourth to fifth columns, and sixth to seventh columns correspond to the $Q_{\alpha}$ values and the logarithmic forms of the $\alpha$ decay half-lives for the three models, respectively. To visually illustrate the impact of the three mass models on half-life calculations, we present the data from Table 5 in graphical form. Fig 3 presents the predicted half-life curves for isotopes with $Z = 118$ and $Z = 120$. The figure clearly shows that $N = 184$ exhibits a significant shell effect across all three models. Notably, the shell effect at $N = 178$ varies depending on the model employed: specifically, it is more pronounced in the WS4 and FRDM models, whereas it appears relatively vague in the RCHB model.

To further validate the reliability of the predicted half-lives, we employ the classical G-N law as a benchmark test\cite{geiger1911lvii}\cite{geiger1922reichweitemessungen}. Established by Geiger and Nuttall in 1911, this law describes the relationship between the half-life of $\alpha$ decay and the $Q$ value, expressed mathematically as
\begin{equation}
\log_{10}(T_{1/2}) = \frac{a}{\sqrt{Q_{\alpha}}}+b,
\end{equation}
where $a$ and $b$ denote the intercept and slope of the linear fit, respectively. In Fig. 4, we can observe that the BNN calibrated CPPM (including nonlocal corrections) predictions adhere well to the expected G-N linear relationship. 

Furthermore, as shown in Fig. 5, we investigate the correlation between the logarithm of predicted half-lives $\log_{10}(T_{1/2})$, and the negative logarithm of penetration probabilities $-\ln$P. The results reveal a robust linear correlation consistent with the G-N law. These findings indicate that the BNN approach not only significantly reduces the RMSE but also accurately preserves the fundamental physical scaling laws of $\alpha$ decay, thereby confirming that our method captures the essential physics of the decay process.

\section{Summary}
We extend the semiclassical WKB approach within the CPPM by introducing a coordinate-dependent effective mass for the emitted $\alpha$ particle, thereby quantifying the impact of nonlocal effects on $\alpha$ decay half-lives. Furthermore, we develop the BNN framework to capture the intrinsic correlations between the half-life and the relevant physical descriptors, including the mass number $A$, proton number $Z$, decay energy $Q_\alpha$, orbital angular momentum $l$, preformation factor $P_\alpha$, quadrupole deformation $\beta_2$, along with residual $t_k$. The application of the BNN to optimize CPPM, CPPM (nonlocal), and three distinct parameterizations of the UDL(Formula-A, Formula-B, Formula-C), leads to a notable enhancement in $\alpha$ decay half-life prediction accuracy, exceeding $23.25\%$ overall. Moreover, combining the BNN with a non-localized method for the optimization of the CPPM results in a further dramatic improvement, with an accuracy gain of up to $47.77\%$. These results demonstrate the robustness and reliability of the BNN-based approach for $\alpha$ decay modeling. Building on this framework, we predict $\alpha$ decay half-lives for the isotopic chains with $Z=118$ and $Z=120$, unveiling pronounced shell effects and showcasing excellent extrapolation capability. The predictions exhibit a linear Geiger-Nuttall-type correlation between $\log_{10} T$ and $Q_\alpha$, consistent with the correlation between $\log_{10} T$ and the negative logarithm of the barrier penetration probability $-\ln$P thereby lending further credence to the method. Overall, BNN opens an effective and versatile avenue for quantitative descriptions of nuclear decay processes.

\section{Acknowledgements}
This work is supported by Yunnan Provincial Science Foundation Project (No. 202501AT070067), Yunnan Provincial Xing Dian Talent Support Program (Young Talents Special Program, No. XDYC-QNRC-2023-0162), Kunming University Talent Introduction Research Project (No. YJL24019), Yunnan Provincial Department of Education Scientific Research Fund Project (No. 2025Y1055 and 2025Y1042), the Program for Frontier Research Team of Kunming University 2023, and National Natural Science Foundation of China (No. 12063006), the Special Basic Cooperative Research Programs of Yunnan Provincial Undergraduate Universities’ Association (grant NO. 202101BA070001-144).

\bibliographystyle{IEEEbib}
\bibliography{bib}
\end{document}